\documentclass[RNAAS]{aastex62}

\received{May 17, 2018}
\accepted{May 23, 2018}
\submitjournal{RNAAS}

\shorttitle{The quadruply lensed quasar WG0214-2105}
\shortauthors{Agnello}

\begin{document}

\title{WG021416.37-210535.3, a quadruply lensed quasar in three public surveys}

\correspondingauthor{Adriano Agnello}
\email{aagnello@eso.org}

\author[0000-0001-9775-0331]{Adriano Agnello}
\affil{European Southern Observatory,\\
 Karl-Schwarzschild-Strasse 2, 85748 Garching bei Muenchen, Germany}\






\section{Discovery of WG0214-2105}
The Southern Hemisphere has just recently begun to be charted by wide-field surveys, with a sufficient depth and image quality to enable the discovery of strongly lensed quasars.
The quadruply imaged quasar  WG0214-2105 (r.a.=02:14:16.37, dec.=-21:05:35.3) is a previously unknown lens, with `blue' mid-IR colors and high UV deficit, found in the intersection of three survey footprints: the Dark Energy Survey public DR1 (DES, Abbott et al. 2018), The VST-ATLAS (Shanks et al. 2015) and Pan-STARRS (Chambers et al. 2016).
Its discovery relied on high spatial resolution from the Gaia mission (Lindegren et al. 2016) and mid-IR color preselection in the WISE catalog (Wright et al. 2010).
Throughout this paper, magnitudes from WISE are in the Vega system.
In what follows, \textit{singlets} (resp. \textit{multiplets}) are objects that correspond to only one (resp. more than one) \texttt{source} entry in the Gaia-DR2 catalog. 

\begin{figure}[b]
\plotone{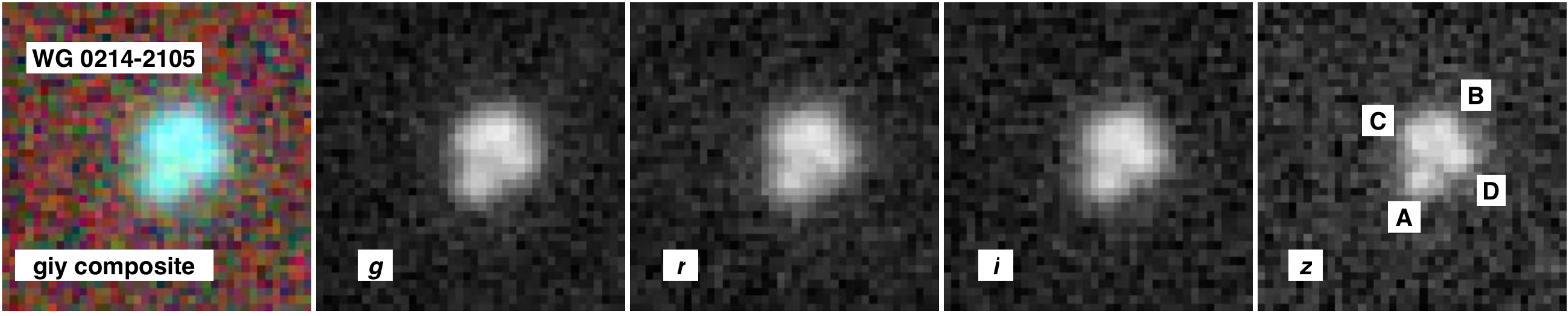}
\caption{Pan-STARRS $giy$ color-composite and $g,r,i,z$ cutouts of WG0214-2105, measuring 10$^{\prime\prime}$ per side; pixels measure $0.258^{\prime\prime}/\mathrm{side}$. The relative displacements $\delta x=-cos(\mathrm{dec.})\delta\mathrm{r.a.}$, $\delta y=\delta\mathrm{dec.}$ of the quasar images with respect to image A, in seconds of arc, are: B=$(0.79,1.53)$, C=$(0.01,1.23)$, D=$(1.16,0.74).$ The maximum image separation is $1.89^{\prime\prime}.$}
\end{figure}

I have used magnitudes from WISE to select $\mathcal{O}(10^6)$ objects with extragalactic colors ($W1-W2>0.2+\sqrt{(\delta W1)^2 +(\delta W2)^2}$) in the Southern Galactic Hemisphere (SGH, $b<-15$).
Upon matching with the Gaia-DR2 catalog, $\mathcal{O}(10^5)$ multiplets are obtained. Many of these are clustered at low $b$ or close to the Magellanic Clouds, and may be line-of-sight alignments of quasars with stars. For this reason, I retained only objects in regions where the density of multiplets is smaller than $0.2$ times the density of singlets, yielding 45549 targets in the SGH, of which 8146 in the 5000~deg$^2$ of the DES footprint.
The DES public cutout service interface was used for visual inspection, simply because it enables the visualisation of multiple cutouts per page.
Besides other known quads, WG021416.37-210535.3 was immediately recognisable due to its `white' optical colors and a characteristic \textit{fold} image configuration.
The image separation and configuration make this system particularly amenable to multi-epoch follow-up, in order to measure the time-delays among light-curves of different images (see e.g. Bonvin et al. 2017 and references therein)
 and hence measure cosmological parameters, primarily the Hubble constant (Refsdal 1964).
Figure 1 shows a $10^{\prime\prime}\times10^{\prime\prime}$ cutout from Pan-STARRS, with the quasar images ordered following the expected arrival times.
The lens galaxy is faint, and follow-up modeling of the image cutouts (beyond the scope of this report) is required in order to obtain accurate relative astrometry and robust lens parameters.

\section{Discussion}
The chosen overdensity threshold is solely empirical, and we can expect more lenses to be found if it is relaxed further. Known quadruplets in the SGH are recovered if the overdensity threshold is raised at $0.29,$ to the price of more onerous visual inspection. These include: HE~0435-12 (Wisotzki et al. 2002), WFI~2033-47 (Morgan et al. 2004), DES~0408-53 (Lin et al. 2017), WGD~2038-40 (Agnello et al. 2017), DES~0405-33 (Anguita et al., subm.).

WG~J021416.37-210535.3 could have been found already in Gaia-DR1, as it is a multiplet in both releases.
Its `blue' $W1-W2=0.42$ WISE color is the reason why it was not found in the WISE-Gaia-DES search (Agnello et al. 2017), which had a more demanding threshold of $W1-W2>0.55$ at preselection.

This system is undetectable in VST-ATLAS $u-$band images (P.~L.~Schechter, private comm.), which suggests a redshift of $z_{s}\gtrsim2.7$ for the source quasar.
Lenses with similar colors but only two quasar images may be easily interpreted, at visual-inspection stage, as line-of-sight pairs of white dwarfs and hence discarded. The fact that other successful searches, relying on population-mixture models to bypass color preselection (Lemon et al. 2018), did not find this system earlier is somewhat surprising.
This discovery then demonstrates the need for accurate classification in wide-field surveys, with special emphasis on rare objects, despite the tremendous improvement on lens searches produced by the Gaia mission.


\acknowledgments

I wish to thank Paul~L.~Schechter and Tommaso Treu for support.\\
This research has made use of the NASA/ IPAC Infrared Science Archive, which is operated by the Jet Propulsion Laboratory, California Institute of Technology, under contract with the National Aeronautics and Space Administration. This research made use of the cross-match service provided by CDS, Strasbourg. This project used public archival data from the Dark Energy Survey (DES).

%

\vspace{5mm}
\facilities{IRSA (WISE, Gaia), CDS Xmatch}

\end{document}